\documentstyle[psfig,epsf]{l-aa}
\def\lsim{\lower.5ex\hbox{$\; \buildrel < \over \sim \;$}}
\def\gsim{\lower.5ex\hbox{$\; \buildrel > \over \sim \;$}}
\newcommand{\eqb}{\begin{eqnarray}}
\newcommand{\eqe}{\end{eqnarray}}

\newbox\grsign \setbox\grsign=\hbox{$>$} \newdimen\grdimen \grdimen=\ht\grsign
\newbox\simlessbox \newbox\simgreatbox
\setbox\simgreatbox=\hbox{\raise.5ex\hbox{$>$}\llap
   {\lower.5ex\hbox{$\sim$}}}\ht1=\grdimen\dp1=0pt
\setbox\simlessbox=\hbox{\raise.5ex\hbox{$<$}\llap
   {\lower.5ex\hbox{$\sim$}}}\ht2=\grdimen\dp2=0pt
\def\gsim{\mathrel{\copy\simgreatbox}}
\def\lsim{\mathrel{\copy\simlessbox}}

\def\ref{\par\noindent\hangindent=1.5cm}
\begin{document}
\thesaurus{}
 
\title{Can DNA bases be produced during molecular cloud collapse?}
 
\author{Sonali Chakrabarti and Sandip K. Chakrabarti}
\institute{S.N. Bose National Centre for Basic Sciences,
JD-Block, Sector III, Salt Lake, Calcutta 700091, India}

\date{Received \dots ; accepted December 23rd, 1999}
\maketitle
\markboth{}{}

\maketitle

\begin{abstract}
Using a reasonably large chemical reaction network consisting
of $421$ species, we show that along with normal chemical evolution
of molecular cloud during collapse
and star formation, significant amount of adenine, a DNA base,
may be produced after an evolution of $10^{6-7}$ years.
This findings may shed light on whether 
life on earth had to begin from scratch or these molecules could 
have contaminated the earth from the beginning.
\end{abstract}

\section{Introduction}

Understanding origin of life on earth is a very challenging astrophysical problem of today.
Study of isotopic ratios on carbon inclusions in ferric salt deposits in Greenland
shows that life existed on earth before $3.85$ billion years
(Mojzsis et al. 1996; Holland 1997; Schidlowski 1993) or possibly even before $4$
billion years (Mukhin \& Gerasimov 1993). On the other hand, 
data from isotope systematics show that
the earth is between $4.55$ to $4.57$ billion years old (Faure 1986). 
The time available for life to develop on the earth from scratch
is possibly around $0.5-0.6$ billion years which is very short (Crick \& Orgel 1973).
One of the ways the formation of life on earth could be assisted if the building blocks of 
life, namely, amino acids and at least some of the bases of DNA could be produced 
well before the planetary formation, presumably during chemical evolution of 
molecular clouds. Whereas extensive work is present in the literature to study evolution of some
of the organic and inorganic molecules, no work is reported so far which study the
formation of bio-molecules or pre-biotic molecules
such as amino acids, sugars, DNA, RNA  etc.

Bases of DNA such as adenine ($H_5C_5N_5$) can be produced by
successive addition of $HCN$ molecules in four steps (Volkenshtein, 1983). In the present 
{\it letter} we perform a hydrodynamic simulation and followed chemical evolution during 
the molecular cloud collapse and show that a significant amount of adenine 
(with mass fraction $X_{adenine}\sim 10^{-10}$) 
could be produced. This indicates that planets and comets could have 
formed with contaminations of pre-biotic molecules. 
If correct, this may resolve a long-standing problem of 
origin of lives on earth.

\section{Reaction Network}
We take the UMIST database (Millar, Farquhar \& Willacy
1997; hereafter referred to as MFW97) as our basis of chemical
reactants and reactions, but added several new reactions such as synthesis of amino acids
(alanine and glycine), hydroxy-acids (glycolic and lactic acids),
DNA base (adenine, see, Volkenshtein 1983), urea synthesis etc. 
These new reactions make the total number of species to be $421$.
The rate co-efficients of these additional reactions are difficult to find, 
especially in the environs of a molecular cloud. To use UMIST database, the rate
constant for a two body reaction is written as (MFW97),
$$
k=\alpha (T/300)^\beta {\rm exp}(-\gamma/T) \ \ {\rm cm}^3\ s^{-1}
\eqno{(1)}
$$
where, $\alpha$, $\beta$ and $\gamma$ are constants and $T$ is the temperature.
Amino acid synthesis rate was estimated from Fig. 8 of Schulte \& Shock (1995).
Urea synthesis rate is kept comparable to the rates given 
in UMIST table. The rate constants were taken to be $\alpha=10^{-10}$, 
$\beta=\gamma=0$ for each two-body reactions. The rate constants for adenine 
synthesis was chosen to be similar to other two body reactions [$\alpha=
10^{-10}$ $\beta=\gamma=0$ for each HCN addition in the chain
$HCN \rightarrow CH(NH)CN \rightarrow NH_2CH(CN)_2 \rightarrow
NH_2(CN)C=C(CN)NH_2 \rightarrow H_5C_5N_5$ (adenine)]. 
This should provide a sufficiently good estimate for the final 
abundance of these pre-biotic molecules.

Initial composition of the cloud before the
simulation begins is kept to be the same as in MFW97, and formation 
of $H_2$ is included using the grain-surface reaction with rates as in MFW97.
The initial {\it mass fractions} are taken to be the same as in
MFW97 (but converted to mass fractions), i.e., H:He:C:N:O:Na:Mg:Si:P:S:Cl:Fe = $0.64$:$0.35897$:$5.6\times 
10^{-4}$:$1.9\times 10^{-4}$:$1.81\times 10^{-3}$:$2.96\times 10^{-8}$:
$4.63\times 10^{-8}$:$5.4 \times 10^{-8}$:$5.79\times10^{-8}$:$4.12\times
10^{-7}$:$9\times 10^{-8}$:$1.08\times10^{-8}$.

\section{Hydrodynamic considerations}

We use simple molecular cloud model with generic initial conditions: initial 
size of the molecular cloud $r_0=3\times 10^{18}$cm, 
average temperature of the cloud $T=10$K, and angular velocity of the cloud $\Omega=10^{-16}$ rad s$^{-1}$. 
The sound speed becomes $a_s=(kT/\mu m_H)^{1/2} \sim 19200$cm s$^{-1}$
and corresponding initial density (Shu, Adams \& Lizano 1987) is $\rho=a_s^2/2\pi G r^2=10^{-22}$g cm$^{-3}$ and 
accretion rate is ${\dot M}=1.06 \times 10^{20}$g s$^{-1}$. Here, $\mu=2.3$ mean 
molecular weight, $m_H$ is the mass of the hydrogen atom, and $k$ is the Boltzmann constant. 
In the isothermal phase of the cloud collapse, density $\rho \propto r^{-2}$ (Chandrasekhar 1939) and the 
velocity is constant. When opacity becomes high enough to trap radiations (say, at $r=r_{tr}$),
the cloud collapses adiabatically with $\rho \propto r^{-3/2}$. In presence of rotation, 
centrifugal barrier forms at $r=r_c$, where centrifugal force balances gravity.
Density falls off as $\rho \propto r^{-1/2}$ in this region  (Hartmann, 1998).
Following Shu, Adams \& Lizano (1987), we compute the density, temperature and velocity distribution
inside the cloud and follow the chemical evolution at the same time.
With density chosen as above, the initial constant velocity of infall becomes $8900$cm s$^{-1}$ and below $r=r_c$
velocity $\propto r^{-1/2}$ to preserve the accretion rate in a disk like structure of constant height.
Since for the parameters chosen (generic as they are) $r_c>r_{tr}$, 
we chose $T\propto 1/r$ inside the centrifugal barrier ($r<r_c$) as in an adiabatic flow. We follow the
collapse till a radius of $10^{12}$cm is reached. Figure 1 shows the temperature and mass density (in units of $10^{20}$)
distribution in a log-log scale. Our choice of $\Omega$ is small by at least an order of 
magnitude from that of the observed value in some of the clouds. This was done to 
obtain atmosphere at a distance of around $1$AU from the proto-star to be as close to reducing ($C/O>1$).

\begin {figure}
\vbox{
\vskip -3.0cm
\hskip 0.0cm
\centerline{
\psfig{figure=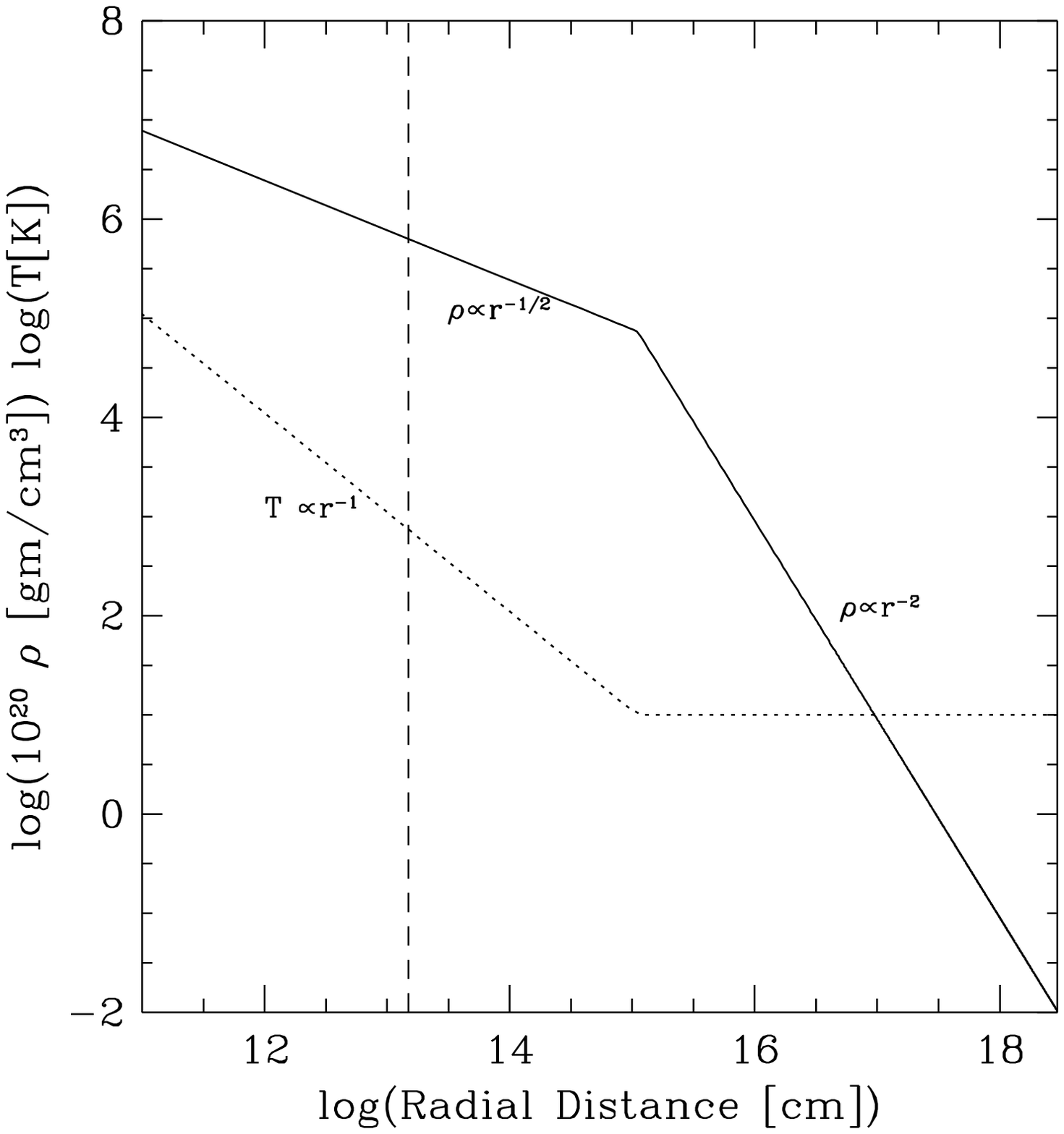,height=10truecm,width=10truecm,angle=0}}}
\vspace{-0.0cm}
\end{figure}
\begin{figure}
\noindent {\small {\bf Fig. 1} : Log-log plot of temperature (dotted) and mass density (in units of $10^{20}$)
as functions of the radial distance. Vertical dashed line is drawn at $1$AU.}
\end{figure}

\section{Results}

Figure 2a shows the log-log variation of mass fractions of some of the light and intermediate 
species of our simulation with radial distance (in cm). For clarity, we 
plot the curves alternately by solid and dotted type. In Fig. 2b, we plot the variation of 
more complex molecules. Since they start with zero abundances, we plot 
them from radius $2.5 \times 10^{18}$cm for clarity. We put a long dashed vertical line at $1$AU, the
distance of the earth with respect to the sun. On the upper axis, we have put time (in seconds) elapsed since the
beginning of collapse at radial distances of $10^{18}$cm, $10^{17}$cm, $10^{16}$cm and $10^{15}$cm respectively. 
Towards the end of the collapse, time spent is negligible.
\begin {figure}
\vbox{
\vskip -4.5cm
\hskip 0.0cm
\centerline{
\psfig{figure=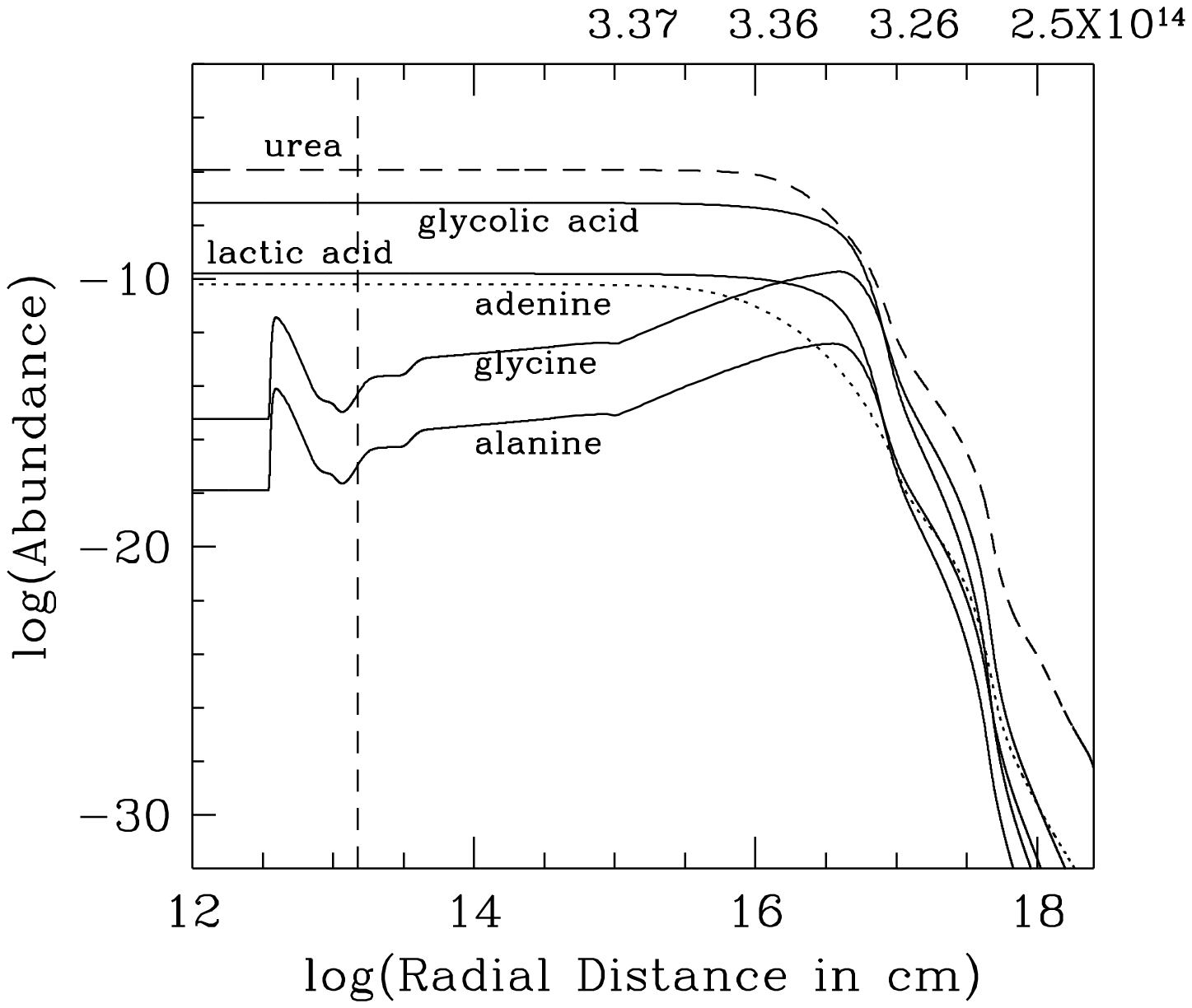,height=12truecm,width=12truecm,angle=0}}}
\vspace{-1.5cm}
\end{figure}
\begin{figure}
\noindent {\small {\bf Fig. 2a} : Log-log plot of the mass fractions of some of the lighter and intermediate 
mass species as functions of the radial distance. Alternate species have been plotted with dotted 
curves for clarity. Upper axis shows time elapsed in seconds since collapse began. 
Vertical dashed line is drawn at 1AU. }
\end{figure}
\begin {figure}
\vbox{
\vskip -5.0cm
\hskip 0.0cm
\centerline{
\psfig{figure=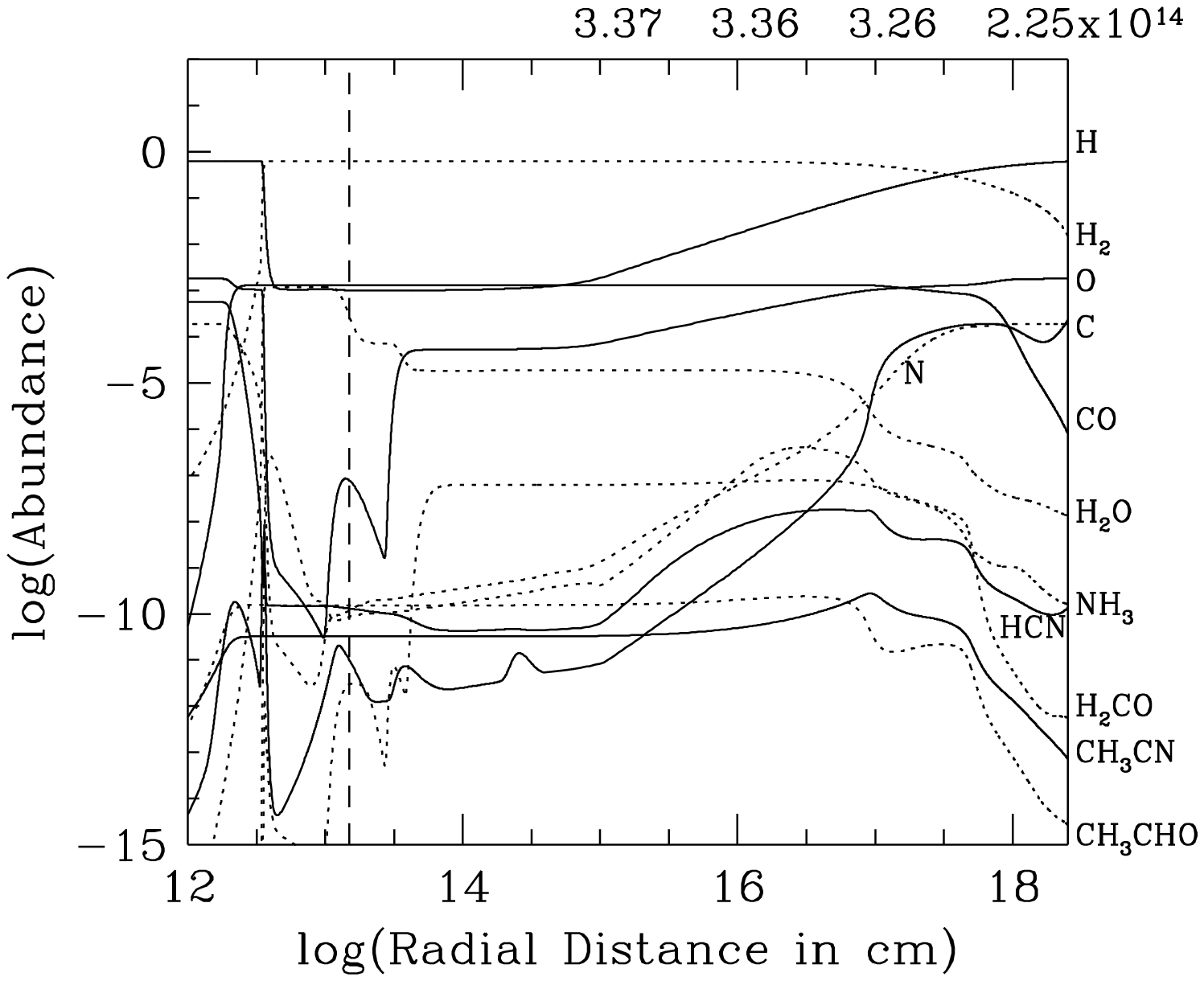,height=12truecm,width=12truecm,angle=0}}}
\vspace{-1.5cm}
\end{figure}
\begin{figure}
\noindent {\small {\bf Fig. 2b} : Log-log plot of the mass fractions of a few complex  molecules (marked)
as functions of the radial distance. Upper axis shows time elapsed in seconds since collapse began.
Vertical dashed line is drawn at 1AU. }
\end{figure}
Because of low initial velocity, $H$, $N$, $C$ and $O$ are depleted much before $100$AU
and $HCN$, $NH_3$, $CH_3CN$ etc. rises as 
also the more complex molecules (Fig. 2b)
which form out of them.  Inside $r_c$, as matter falls faster and spends lesser
time, the depletion of lighter molecules are controlled, until the density and temperature
also rises so high that depletion started once more. Inside, $3 \times 10^{12}$ cm,
the composition changes in a very short timescale
Around $r\sim 10$AU, the mass fractions of adenine, urea and glycine
are already significant. At $r=1$AU, $X_{adenine}= 6 \times 10^{-11}$.
Since the mass of the earth is around $4\times 10^{27}$g, this corresponds to
$2.4 \times 10^{17}$g of adenine which could have contaminated the earth
at the time of formation (it could be higher since metallic content of
earth is much above the average molecular value. Also, one has to include the
effect of dust-gas mixture in the molecular cloud) This computation, does not 
consider the destructions of adenine at a higher temperature 
region, and it is likely that much of these contaminants are destroyed during
collapse and formation of proto-earth. However, comets formed in the inner cloud 
could carry away these pre-biotic molecules and deposit them during 
future impacts on planets. It is to be noted that  around $1$AU, the composition is 
close to reducing ($C/O>1$) type (Fig. 2a) which is favourable for the formation of bio-molecules.

Since regions of low density of molecular cloud could have a very long evolution time scale,
it may of interest whether pre-biotic molecules could have formed in a very low density 
isothermal region. To this effect, we choose a cloud of mass density $\rho=1.6 \times 10^{-19}$g cm$^{-3}$
and $T=10$K and let it evolve for $10^6$ years.  Figure 3 shows the results of this
simulation where we plot the variation of abundance with time (in seconds). 
We find that the abundance of adenine, for instance, is around $5 \times 10^{-12}$. This is
a very generic condition, and the abundance is significant. We therefore believe that
adenine could be produced during the molecular cloud collapse.

\begin {figure}
\vbox{
\vskip -3.5cm
\hskip 0.0cm
\centerline{
\psfig{figure=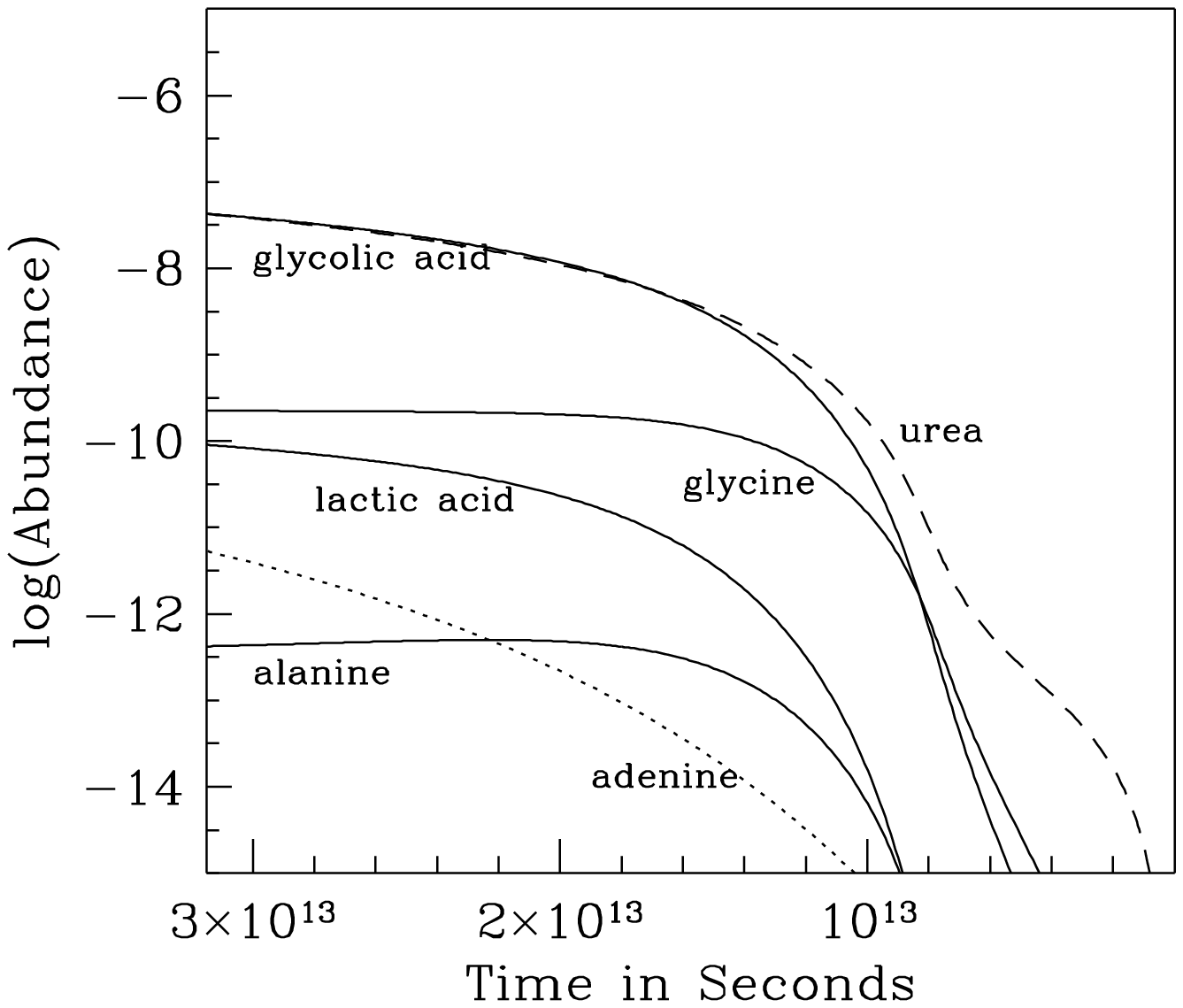,height=12truecm,width=12truecm,angle=0}}}
\vspace{-2.5cm}
\end{figure}
\begin{figure}
\noindent {\small {\bf Fig. 3}: 
Log-log plot of the mass fractions of some pre-biotic molecules (marked)
as functions of time during the evolution of a static cloud of constant density and temperature 
for $10^6$ years. }
\end{figure}

\section{Concluding remarks}

It is generally thought that molecules vital to living systems on earth may have started from scratch.
In this {\it Letter}, we have explored the possibility of formation of a DNA base, namely adenine,
and a few of the simpler amino acids during a generic molecular cloud collapse and show that mass fraction of adenine
is significant which might have contaminated the planetary disk even before the
formation of the planets themselves. In a static evolution of a cloud of constant density and 
temperature, the formation of adenine is also important, and therefore our conclusion may be very generic.
The rates we have chosen are reasonable, though uncertainties 
of an order of magnitude cannot be ruled out at this stage as more accurate estimates
require one to have temperature dependent rate coefficients of both forward and reverse reactions.

Since earth is a generic planet and since it seems apparent that bio-molecules are as naturally produced as
any other complex molecules, we believe that DNA bases produced in collapsing cloud could 
have contaminated the earth, and at the same time there should be many such planets in each galaxy where
DNA based lives should flourish. It is not yet clear whether or not these molecules survived during further
evolution of planetary disks leading to eventual formation of planets
or actual contaminations were made much later by comets carrying away protecting  these molecules.
These vital issues will be addressed  in due course.

\acknowledgements

Authors thank Tom Millar for sending UMIST database electronically
and the referee for several helpful comments. 
SC thanks the Council for Scientific and Industrial Research (CSIR)
for financial support and SNBNCBS for technical support.  

{}

\end{document}